\newcommand*{\ATLASLATEXPATH}{}
\documentclass[PAPER, atlasdraft=false, cernpreprint=true, texlive=2016, USenglish, txfonts]{\ATLASLATEXPATH atlasdoc}
 
\usepackage{\ATLASLATEXPATH atlaspackage}
\usepackage{\ATLASLATEXPATH atlasbiblatex}
 
\usepackage{\ATLASLATEXPATH atlascontribute}
 
\usepackage{\ATLASLATEXPATH atlasphysics}
\usepackage{\ATLASLATEXPATH atlasheavyion}
\usepackage{verbatim}
\usepackage{rotating}
 
\graphicspath{{logos/}{figures/}}
 
\AtlasTitle{Measurement of azimuthal anisotropy of muons from charm and bottom hadrons in $pp$ collisions at $\sqrt{s}=13$ TeV with the ATLAS detector}
 
\AtlasRefCode{HION-2018-23}
 
\AtlasNote{HION-2018-23}
 
\AtlasJournalRef{Phys.\ Rev.\ Lett. 124 (2020) 082301}
\AtlasDOI{10.1103/PhysRevLett.124.082301}
 
\AtlasAbstract{
The elliptic flow of muons from the decay of charm and bottom hadrons
is measured in $pp$ collisions at $\sqrt{s}=13$ TeV using a data
sample with an integrated luminosity of 150~pb$^{-1}$ recorded by the
ATLAS detector at the LHC.  The muons from heavy-flavor decay are
separated from light-hadron decay muons using momentum imbalance
between the tracking and muon spectrometers.  The heavy-flavor decay
muons are further separated into those from charm decay and those from
bottom decay using the distance-of-closest-approach to the collision
vertex.  The measurement is performed for muons in the transverse
momentum range 4--7~GeV and pseudorapidity range $|\eta|<2.4$.  A
significant nonzero elliptic anisotropy coefficient $v_{2}$ is
observed for muons from charm decays, while the $v_{2}$ value for
muons from bottom decays is consistent with zero within uncertainties.
}

\PreprintIdNumber{CERN-EP-2019-166}
\hypersetup{pdftitle={ATLAS document},pdfauthor={The ATLAS Collaboration}}
 
\newcommand{\dphi}{\mbox{$\Delta\phi$}\xspace}
\newcommand{\deta}{\mbox{$\Delta\eta$}\xspace}
\newcommand{\Nchrec}{\mbox{$N_\mathrm{ch}^\mathrm{rec}$}\xspace}
\newcommand{\dca}{\mbox{$d_{0}$}\xspace}
\newcommand{\fsig}{\mbox{$f^\mathrm{sig}$}\xspace}
\newcommand{\fbmu}{\mbox{$f^{b\to\mu}$}\xspace}
\newcommand{\fcmu}{\mbox{$f^{c\to\mu}$}\xspace}
\newcommand{\pimbal}{\mbox{$\Delta p / p_\mathrm{ID}$}\xspace}
\newcommand{\vttsig}{\mbox{$v_{2,2}^\mathrm{sig}(\pt^{\mu}, \pt^{h})$}\xspace}
\newcommand{\vttbkg}{\mbox{$v_{2,2}^\mathrm{bkg}(\pt^{\mu}, \pt^{h})$}\xspace}
\newcommand{\vttcmu}{\mbox{$v_{2,2}^{c}(\pt^{\mu}, \pt^{h})$}\xspace}
\newcommand{\vttbmu}{\mbox{$v_{2,2}^{b}(\pt^{\mu}, \pt^{h})$}\xspace}
 
\newcolumntype{C}[1]{>{\centering\arraybackslash}p{#1}}

\begin{document}
 
\maketitle
 
In high-energy collisions between large nuclei at the Relativistic
Heavy Ion Collider (RHIC) and the Large Hadron Collider (LHC), a
quark--gluon plasma is formed, which rapidly expands as described by
nearly inviscid hydrodynamics~\cite{Romatschke:2017ejr,Heinz:2013th}.
Heavy quarks, which have a large mass and dead-cone radiation
region~\cite{Dokshitzer:2001zm}, are expected to interact with the
medium through a different interplay of radiative and collisional
processes with respect to ordinary light quarks~\cite{Sharma:2009hn}.
However, it was hypothesized that even these massive heavy quarks may
scatter within the medium and be redirected in a way that results in
collective flow patterns~\cite{Batsouli:2002qf}.  Measurements of
decay electrons from charm and bottom hadrons by the PHENIX experiment
in \AuAu collisions at a nucleon--nucleon center-of-mass energy
$\sqrt{s_\textrm{NN}}= 200$~\GeV\ revealed that heavy quarks undergo
significant scattering in the medium and thus lose energy and align
with the geometry of the expanding medium~\cite{Adare:2006nq}.  More
recent measurements using decay leptons and full reconstruction of
charm and bottom hadrons indicate substantial modifications to the
momentum distributions of heavy quarks in heavy-ion collisions
relative to that in proton--proton (\pp) collisions at both RHIC and
the LHC (see Ref.~\cite{Dong:2019byy} for a recent review).
 
Smaller collision systems, including \pPb and even \pp, have particle
emission patterns with large azimuthal anisotropies, also described by
nearly inviscid
hydrodynamics~\cite{Khachatryan:2010gv,Romatschke:2017ejr,Nagle:2018nvi}.
A common hydrodynamic description of \pp, \pPb, and \PbPb azimuthal
anisotropies as resulting from initial geometry anisotropies is
compelling~\cite{Weller:2017tsr}.  New measurements of similar
anisotropies for reconstructed $D$ mesons and heavy-flavor decay
electrons in \pPb collisions~\cite{Sirunyan:2018toe,Acharya:2018dxy}
highlight that charm quarks are scattered in the medium in smaller
collision systems as well.  These measurements of anisotropies with
almost no modification to the transverse momentum (\pt)
distribution~\cite{Adam:2016ich} are somewhat surprising, because such
scattering in the \NucNuc case leads simultaneously to azimuthal
anisotropies and a softening of the transverse momentum
distributions~\cite{Moore:2004tg}.  It is of interest to measure
heavy-flavor anisotropies in \pp collisions in order to obtain
information about the interaction of heavy quarks with the medium in
the smallest hadronic collision system at the LHC.  In this Letter,
measurements of azimuthal anisotropies for muons from heavy-flavor
decays in \pp collisions at 13~\TeV\ are presented.  Additionally, the
heavy-flavor muons are separated to provide information about the
anisotropies of muons from charm and bottom decay separately.

The ATLAS experiment~\cite{PERF-2007-01} is a multipurpose particle
detector with a forward--backward symmetric cylindrical geometry and a
near \(4\pi\) coverage in solid angle.
[Coordinate system with its origin at the nominal interaction point
(IP) in the center of the detector and the \(z\) axis along the beam
pipe.  The \(x\) axis points from the IP to the center of the LHC
ring, and the \(y\) axis points upwards.  Cylindrical coordinates
\((r,\phi)\) are used in the transverse plane, \(\phi\) being the
azimuthal angle around the \(z\) axis.  The pseudorapidity is
defined in terms of the polar angle \(\theta\) as \(\eta = -\ln
\tan(\theta/2)\).  Angular distance is measured in units of \(\Delta
R \equiv \sqrt{(\Delta\eta)^{2} + (\Delta\phi)^{2}}\).
]
It consists of an inner tracking detector surrounded by a thin
superconducting solenoid providing a \SI{2}{\tesla} axial magnetic
field, electromagnetic and hadron calorimeters, and a muon
spectrometer.  The inner tracking detector (ID) covers the
pseudorapidity range \(|\eta| < 2.5\).  It consists of silicon pixel,
silicon microstrip, and transition radiation tracking detectors.  The
muon spectrometer (MS) surrounds the calorimeters and is based on
three large air-core toroidal superconducting magnets with eight coils
each.  The muon spectrometer includes a system of precision tracking
chambers and fast detectors for triggering. The trigger system
consists of a hardware-based first-level trigger and a software-based
high-level trigger (HLT)~\cite{TRIG-2016-01}, which reconstructs the
event in a manner similar to that performed offline.
 
Data for this analysis were recorded during a special running period
in 2017 in which the mean number of \pp interactions per beam crossing
was two.
Events were recorded using triggers that
require a muon at the HLT stage with \pt larger than 4~\GeV~in
coincidence with various triggers designed to select high-multiplicity
events~\cite{HION-2016-01}.
The latter included requirements on the
transverse energy in the calorimeter, and the number of space points
recorded in the silicon microstrip detector, and the number of
charged-particle tracks reconstructed by the HLT.
The trigger with the highest threshold for the number of tracks sampled $150$~pb$^{-1}$, while triggers with lower thresholds were prescaled and sampled less integrated luminosities.
For each charged-track multiplicity range reported here, analyzed events are taken
exclusively from the trigger that sampled the largest integrated
luminosity.

Charged-particle tracks and collision vertices are reconstructed in
the ID using algorithms described in
Ref.~\cite{ATL-PHYS-PUB-2015-018}.  Tracks with $\pt>0.4$~\GeV\ and
$|\eta|<2.5$ satisfying the set of quality
requirements~\cite{HION-2016-01} are used in this analysis.  Muons
with $4<\pt<7$~GeV\ and $|\eta|<2.4$ reconstructed in both the ID and
the MS are selected and required to pass ``medium'' selection
requirements described in Ref.~\cite{PERF-2015-10}.  Events are
required to have at least one but not more than four reconstructed
vertices to reduce the contribution from in-time pileup events
containing multiple \pp collisions per event.  The number of
reconstructed tracks with $\pt>0.4$~\GeV\ associated with the vertex
containing the muon is denoted by \Nchrec.
 
Simulated events were generated using
\PYTHIA{8}~\cite{Sjostrand:2007gs} with the NNPDF23LO parton
distribution function set~\cite{Ball:2012cx} and
A14~\cite{ATL-PHYS-PUB-2014-021} set of tuned parameters. Multijet
hard-scattering events filtered on the presence of a generator-level
muon were passed through a \textsc{Geant}4
simulation~\cite{Agostinelli:2002hh,SOFT-2010-01} of the detector and
reconstructed under the same conditions as the data including pileup background events.
A muon trigger emulator is included in the simulation to evaluate the trigger efficiency.
 
This analysis follows two-particle correlation methods used in
previous ATLAS measurements~\cite{HION-2015-09,HION-2016-01} and
summarized here.  Two-particle correlations are measured as a function
of $\dphi\equiv\phi^{\mu}-\phi^{h}$ and
$\deta\equiv\eta^{\mu}-\eta^{h}$, where particles $\mu$ and $h$ are
muons and charged hadrons, respectively.  For each muon, correlation
functions $S(\deta,\dphi)$ and $B(\deta,\dphi)$ are
formed~\cite{Adare:2008ae}.  The correlation function $S(\deta,\dphi)$
uses charged hadrons from the same event.  The function
$B(\deta,\dphi)$ is constructed by selecting charged hadrons from
different events of similar \Nchrec ($|\Delta\Nchrec|<10$) and vertex
position \zvtx ($|\Delta\zvtx|<10~\mathrm{mm}$).  Detector acceptance
effects largely cancel out in the ratio $S/B$ within the precision of these measurements.  Each muon--hadron pair
is weighted by the inverse product of the trigger and reconstruction
efficiencies for the muon and the reconstruction efficiency for the
charged hadron.
 
One-dimensional correlation functions $C(\dphi)$ are obtained by
integrating $S(\deta,\dphi)$ and $B(\deta,\dphi)$ over the
pseudorapidity interval $1.5<|\deta|<5$:
 
\begin{equation*}
C(\dphi) = \frac{\int_{1.5}^{5} d|\deta| S(|\deta|,\dphi)}{\int_{1.5}^{5} d|\deta| B(|\deta|,\dphi)} \equiv \frac{S(\dphi)}{B(\dphi)},
\end{equation*}
and $S(\dphi)$ and $B(\dphi)$ are normalized such that the average
value of $C(\dphi)$ is unity. Requiring a gap in $\Delta\eta$ that
excludes |$\Delta\eta|<1.5$ reduces the contribution to the
correlations from jet fragmentation.  Previous hadron--hadron
correlation results used a larger gap, integrating over
$2<\left|\deta\right|<5$ instead~\cite{HION-2015-09,HION-2016-01};
however, studies of shape variation versus different $|\Delta\eta|$ selections with the \PYTHIA{8} sample described above indicate
that the jet-fragmentation correlation for heavy-flavor quarks is insignificant for muon-hadron pairs with |$\Delta\eta|>1.5$.
 
In order to separate the flow contribution from back-to-back dijets and resonance decays, together referred to as
nonflow, a template fitting method developed for previous ATLAS
analyses~\cite{HION-2015-09,HION-2016-01} is used.  This method
assumes that the shape of nonflow correlations is independent of the
particle multiplicity in the events, an assumption which results in a
good description of the correlation functions in these measurements
and is tested in simulation~\cite{Lim:2019cys}.  Hence the correlation
function in low particle-multiplicity (LM) events dominated by
nonflow is used to estimate the nonflow contribution in high
multiplicity (HM) events.  The resulting template fit function:
 
\begin{equation*}
C^\mathrm{templ}(\dphi) = F C^\mathrm{LM}(\dphi) + C^\mathrm{ridge}(\dphi),
\end{equation*}
where
 
\begin{equation*}
C^\mathrm{ridge}(\dphi) = G\bigg[ 1+ \sum_{n=2}^{4} 2 v_{n,n} \cos(n\dphi) \bigg],
\end{equation*}
has free parameters $F$ and $n^\text{th}$-order flow (anisotropy)
coefficients $v_{n,n}$; the coefficient $G$ is fixed by requiring that
the integrals of $C^\mathrm{templ}(\dphi)$ and $C(\dphi)$ are equal.
The template fits include harmonics 2--4 because the contribution from
higher-order coefficients is negligible.  Based on the assumption of
flow factorization~\cite{HION-2011-01}, the flow coefficients $v_n$ of
muons are obtained as $v^{\mu}_n(\pt^{\mu}) =
v_{n,n}(\pt^{\mu},\pt^{h})/v^{h}_{n}(\pt^{h})$, where
$v^{h}_{n}(\pt^{h})$ are the flow coefficients of charged hadrons
previously measured by ATLAS using the same template fit method in
different analyses~\cite{HION-2015-09,HION-2016-01}.
 
The selected muon sample includes background muons from particles
produced from light-hadron decay and from punch-through hadrons.  Previous
studies~\cite{STDM-2011-07,HION-2015-06} showed that the signal
(heavy-flavor) and background muons can be separated statistically
using the fractional momentum imbalance, $\pimbal =
\left(p_\mathrm{ID} - p_\mathrm{MS}\right)/p_\mathrm{ID}$, where
$p_\mathrm{ID}$ is the muon momentum measured in the ID, and
$p_\mathrm{MS}$ is that measured in the MS corrected via simulation
for the energy loss inside the calorimeter.  The signal fraction \fsig
is obtained by fitting the measured $\pimbal$ distribution with signal
and background template distributions obtained from simulation.  The
signal muon sample includes remaining contributions from
non-heavy-flavor components such as quarkonia, low-mass resonances,
and $\tau$ leptons; these amount to $\sim2.5\%$, based on \PYTHIA{8}
simulation.

\begin{figure*}[hbt]
\includegraphics[width=0.49\linewidth]{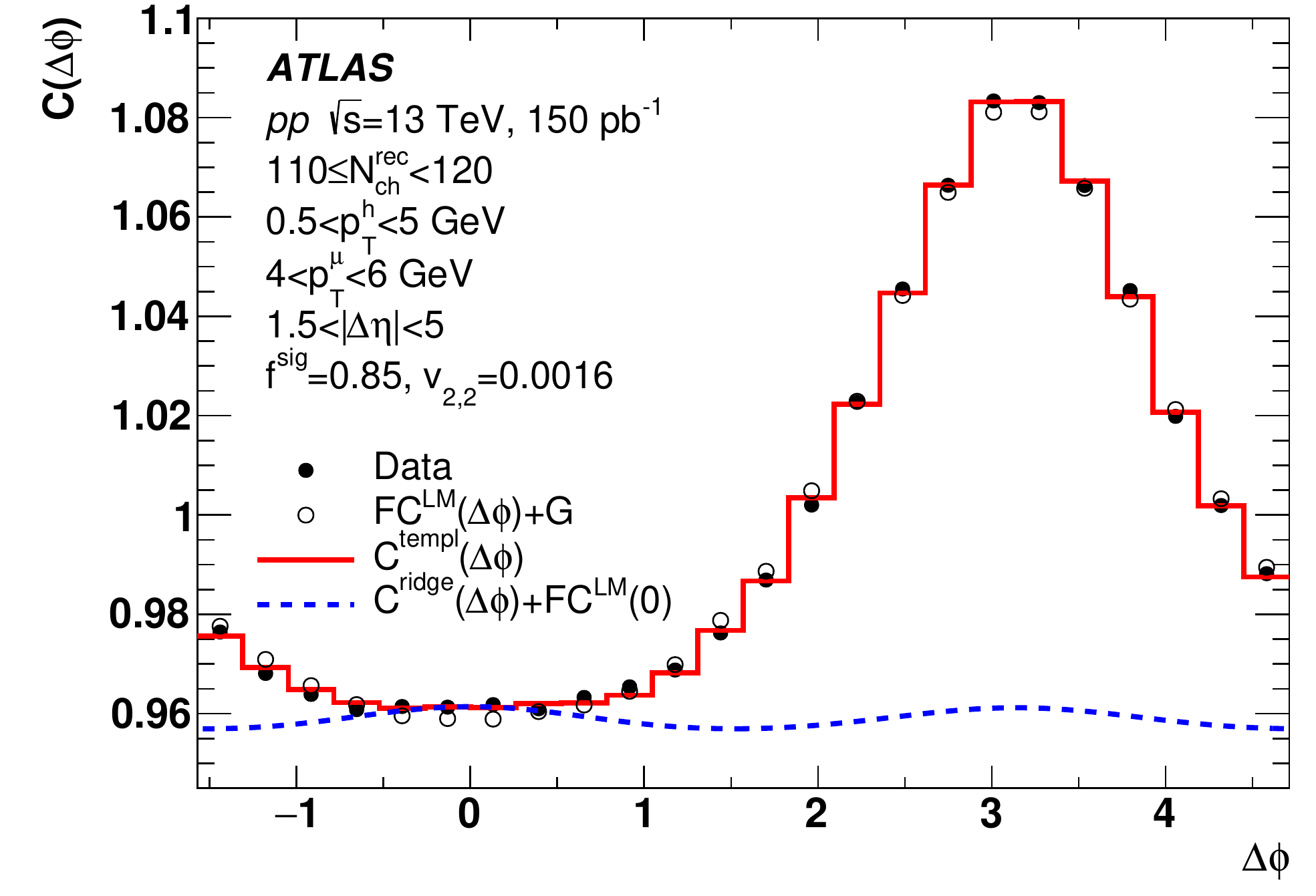}
\includegraphics[width=0.49\linewidth]{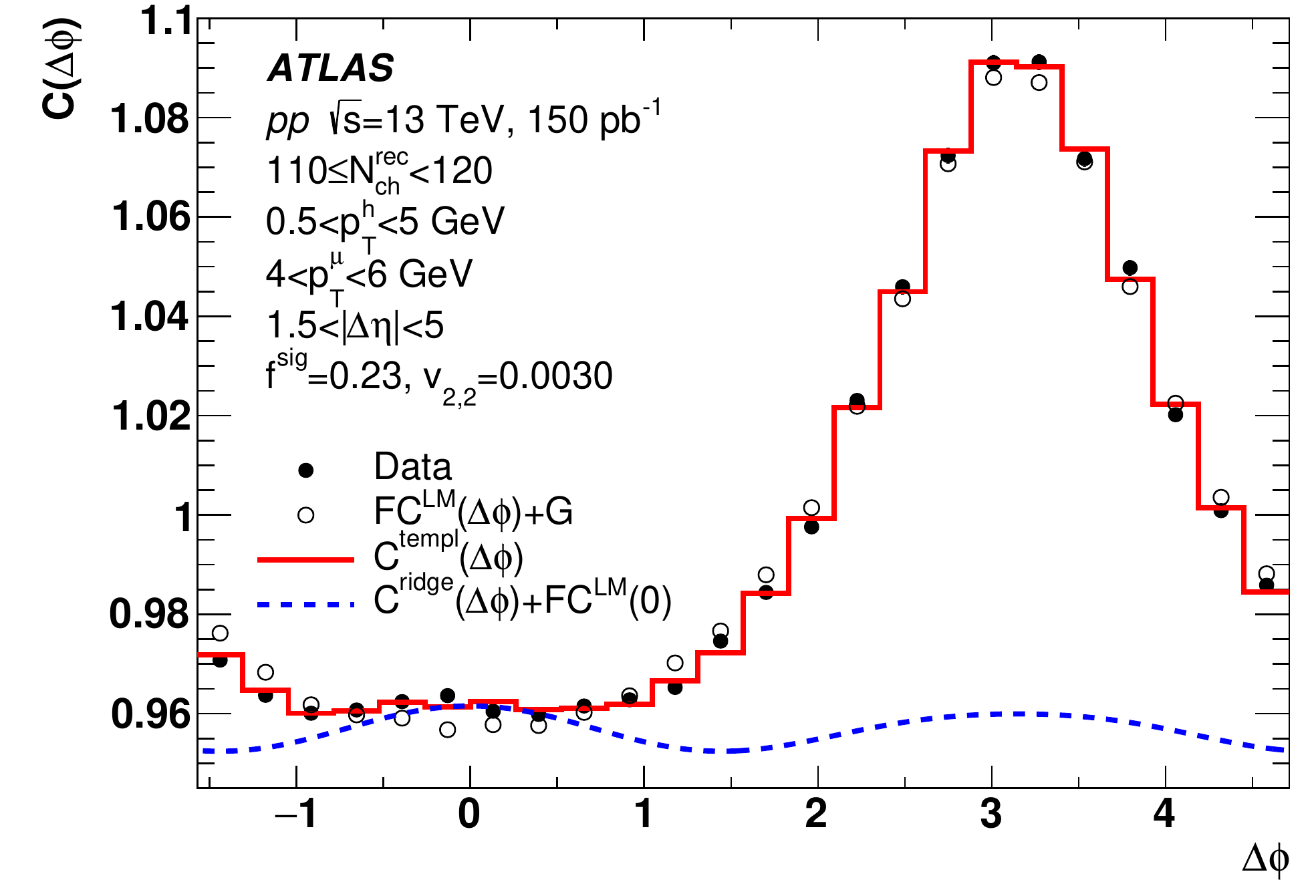}
\caption{\label{fig:cf} Template fit to the muon--hadron correlation
function, $C(\dphi)$, with pseudorapidity interval $1.5<|\deta|<5$
and track multiplicity $110\leq\Nchrec<120$. Muons with transverse
momentum $4<\pt<6$~\GeV\ and charged particles with
$0.5<\pt<5$~\GeV\ are used. Each panel shows the muon--hadron
correlation function for muons of a different signal fraction
(\fsig). The solid red lines show the final function
$C^\mathrm{templ}(\dphi)$, while the open points and dashed blue
lines show the scaled $C^\mathrm{LM}(\dphi)$ and $v_{n,n}$
components, each above a vertical pedestal for visibility.}
\end{figure*}

Figure~\ref{fig:cf} shows muon-hadron correlation functions and
template fits for muons with $4<\pt<6$~\GeV\ and charged hadrons with
$0.5<\pt<5$~\GeV\ from events with $110\leq \Nchrec<120$; the
$\Nchrec<40$ region is used for LM events.  The two panels represent
different \pimbal regions, characterized by different \fsig values, as
indicated in the plots.  The amplitude of the $v_{2,2}$ modulation
changes with the signal fraction.  The values of $v_{2,2}$ are
determined from muon-hadron correlation functions generated using
muons in three different regions of \pimbal, and $v_{2,2}$ as a
function of \fsig is extracted from a linear fit to the points.  Then
$v_{2,2}$ from heavy-flavor muon-hadron correlations \vttsig is
calculated by extrapolating to $\fsig=1$, based on
 
\begin{equation*}
v_{2,2}(\pt^{\mu}, \pt^{h}) = \fsig \vttsig + (1-\fsig) \vttbkg,
\end{equation*}
where \vttbkg is $v_{2,2}$ from background muon-hadron correlations.
 
\begin{figure}[t]
\centering
\includegraphics[width=0.5\linewidth]{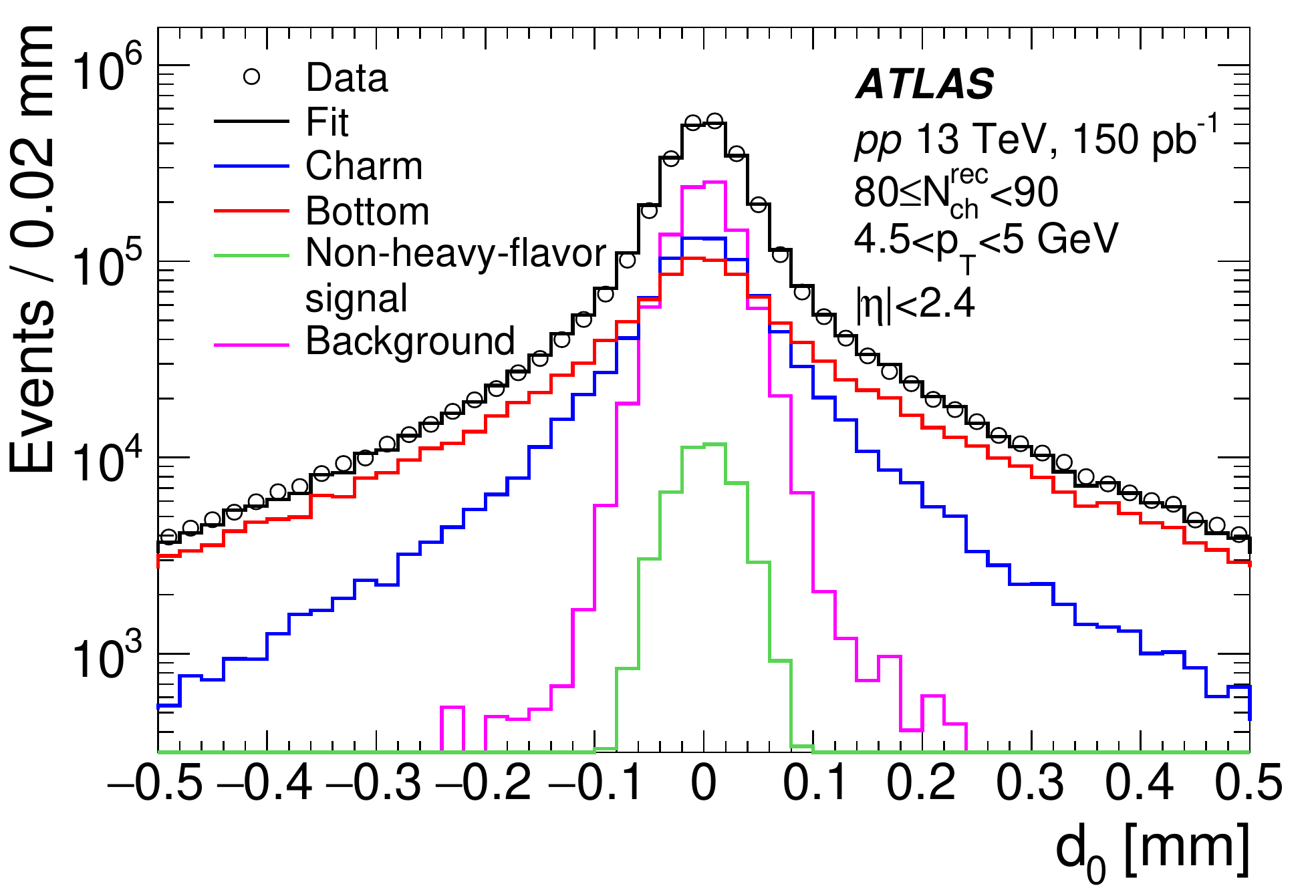}
\caption{\label{fig:dca_fit} Fit (gray histogram) to the transverse
impact parameter, \dca, distribution of muons with transverse
momentum $4.5<\pt<5$~\GeV\ and pseudorapidity $|\eta|<2.4$, with
template \dca distributions of different components obtained from
\PYTHIA{8} simulations.}
\end{figure}
 
Muons from heavy-flavor decays can be further separated into those
from charm and those from bottom decays, based on the different decay
lengths of charm and bottom hadrons.  Template distributions of the
impact parameter of the muon relative to the associated collision
vertex in the transverse direction (\dca) for charm, bottom,
non-heavy-flavor signal, and background muons obtained from the full
detector simulation are used to fit the data distributions
differentially in $p_\mathrm{T}$ and $\eta$.  The \dca resolution of
charged hadrons with $4<\pt<7$~\GeV\ is 20--40~$\mu\mathrm{m}$,
depending on \pt and $\eta$, and independent of
$N_\mathrm{ch}^\mathrm{rec}$. Figure~\ref{fig:dca_fit} shows the fit
to the \dca distribution for muons with $-0.2<\pimbal<0.4$ and $4.5<\pt<5$~\GeV\ in events
with $80 \leq \Nchrec<90$.  The background fraction is fixed in accord
with the fit results in \pimbal.  The contribution from
non-heavy-flavor signal muons is also fixed, using the fraction
obtained from \PYTHIA{8} simulation.  The $|\dca|<0.02~\mathrm{mm}$
region is dominated by non-heavy-flavor signal and background muons and is excluded in the fit procedure.  The fraction of muons from bottom
decays relative to all heavy-flavor muons, $\fbmu=(b\to\mu)/(c\to\mu +
b\to\mu)$, is found to be $\sim0.4$ at $\pt=4$~\GeV\ and increases to $\sim0.6$ at
$\pt=7$~\GeV.  These values are compatible with those determined via a
fixed-order plus next-to-leading-logarithm (FONLL)
calculation~\cite{Cacciari:2012ny} and the \PYTHIA{8} simulation.
 
In order to measure $v_{2,2}$ from charm muon-hadron correlations and
bottom muon-hadron correlations separately, muons are divided into
two \dca regions, $|\dca|<0.12~\mathrm{mm}$ and
$|\dca|>0.12~\mathrm{mm}$.  In the $|\dca|<0.12~\mathrm{mm}$ region
where $\fcmu>\fbmu$, there is a significant hadronic background contribution
and thus \vttsig is obtained in three different \pimbal bins and
extrapolated to $\fsig=1$.  In contrast, in the region
$|\dca|>0.12~\mathrm{mm}$ where $\fcmu<\fbmu$, there is negligible
background and thus \vttsig is obtained directly.  Given two \vttsig
values with different \fbmu values, \vttcmu and \vttbmu can be
determined separately.
 
The sources of systematic uncertainty in \vttsig originate from the LM
event selection, the $\deta$-gap selection, event pileup, trigger and
reconstruction efficiency, and signal fraction extraction.  The impact
on the \vttsig measurement is evaluated by repeating the analysis with
variations intended to test the sensitivity to these effects.  In many
cases, the evaluated variation in \vttsig is driven by statistical
fluctuations.  Sensitivity to the choice of the LM range may arise due
to a change in the dijet shape from the LM to HM events.  The
uncertainty is studied using two alternative $\Nchrec$ ranges, 0--30
and 20--40, for $C^\textrm{LM}(\dphi)$. The resulting variation in
\vttsig is 15--35\% depending on \Nchrec, and is the largest
systematic uncertainty.  The sensitivity to the width of the $\deta$
gap is tested by using $2<|\deta|<5$ to obtain a wider excluded range
($|\deta|<2$), and the resulting change in \vttsig is smaller than the
statistical uncertainty.  The results may be sensitive to a residual
in-time pileup contribution when two closely spaced \pp events are
reconstructed with a single merged vertex.  This effect is studied
using a tighter event selection to reject events containing more than
two reconstructed vertices per event.  The \vttsig obtained is
consistent with the \vttsig from the nominal event selection within
the statistical uncertainties.  The uncertainty associated with the
signal fraction extraction is evaluated by modifying the
momentum-imbalance templates from simulation, and considering the
systematic uncertainties in the muon momentum resolution and
scale. For signal muons, the impact of using a data-driven template
with muons from $J/\psi\to\mu\mu$ candidates is also considered.
No systematic uncertainty on the \vttsig is assigned from this study.
 
The uncertainty in \vttcmu and \vttbmu additionally includes the
uncertainty in \fbmu. The extracted \fbmu values are sensitive to the
shape of the \pt spectra of initial charm and bottom hadrons, the
background muon fraction, the non-heavy-flavor muon contribution, fit
range, and \dca-template shapes.
The shape of the initial hadron \pt distribution is varied from \PYTHIA{8}
to that from a fixed-order plus next-to-leading-logarithm.
The non-heavy-flavor muon
contribution, which is estimated to be 2.5\% of the signal muon yield
using simulation, is varied in the range 0\%--5\% and included in the
\dca fit procedure to evaluate the impact on \fbmu.
The sensitivity to the fit range is evaluated by repeating the \dca fit with different exclusion regions, either 0 or 0.04~mm, and the uncertainty from the \dca-template shape is evaluated with \dca-template shape variation extracted from the \dca shape comparison between the data and simulation.
These variations are included in the final systematic uncertainties.
The resulting systematic uncertainty in \fbmu is 8\%--10\%, and this uncertainty is propagated
into the uncertainties in \vttcmu and \vttbmu by combining it in
quadrature with those in \vttsig.  Finally, it was checked in the
generation-level and reconstruction-level \PYTHIA{8} events that
$v_{2,2}(\pt^{\mu},\pt^{h})$ for inclusive heavy-flavor muons as well
as for muons from $c$ and $b$ decays is consistent with zero as
expected.

\begin{figure*}[hbt]
\includegraphics[width=0.49\linewidth]{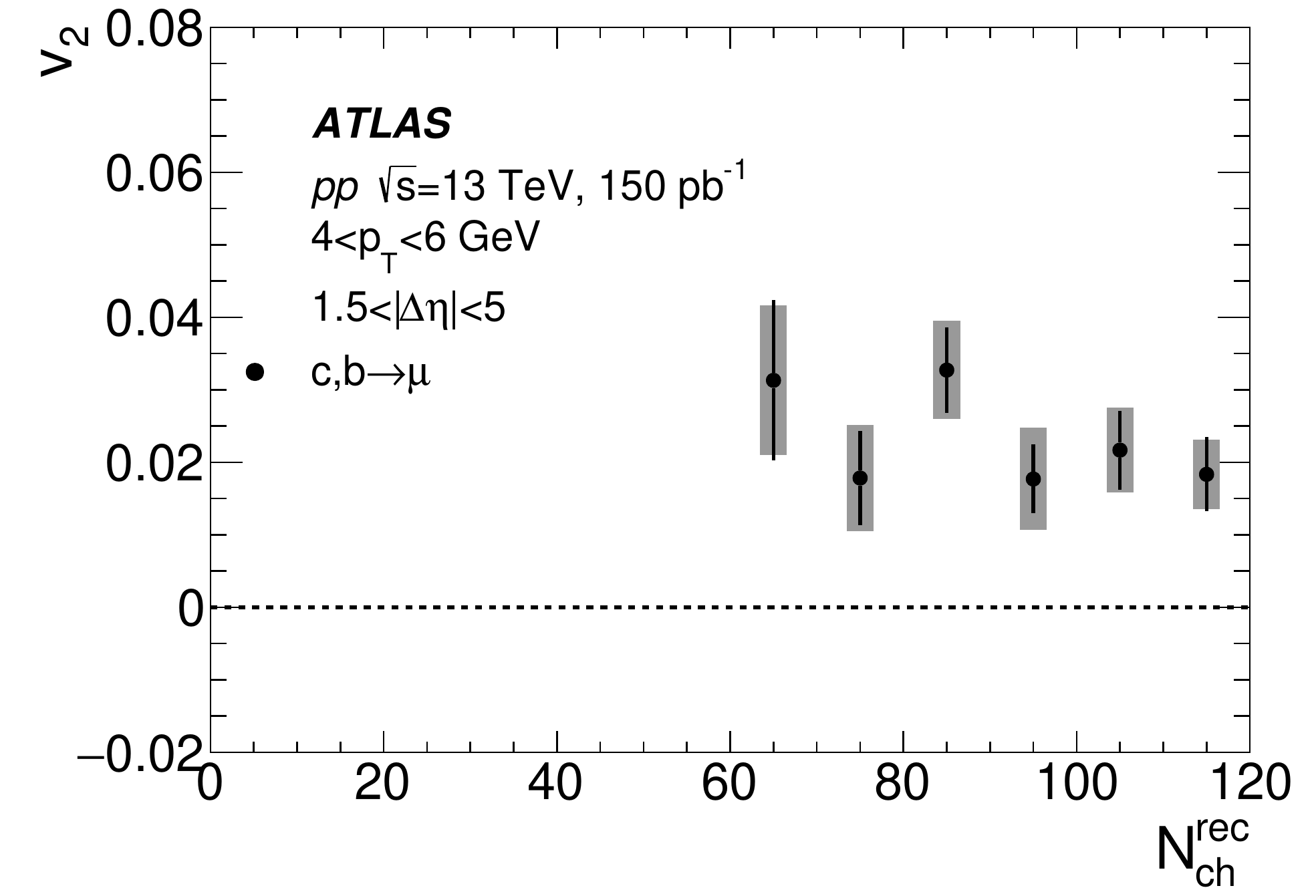}
\includegraphics[width=0.49\linewidth]{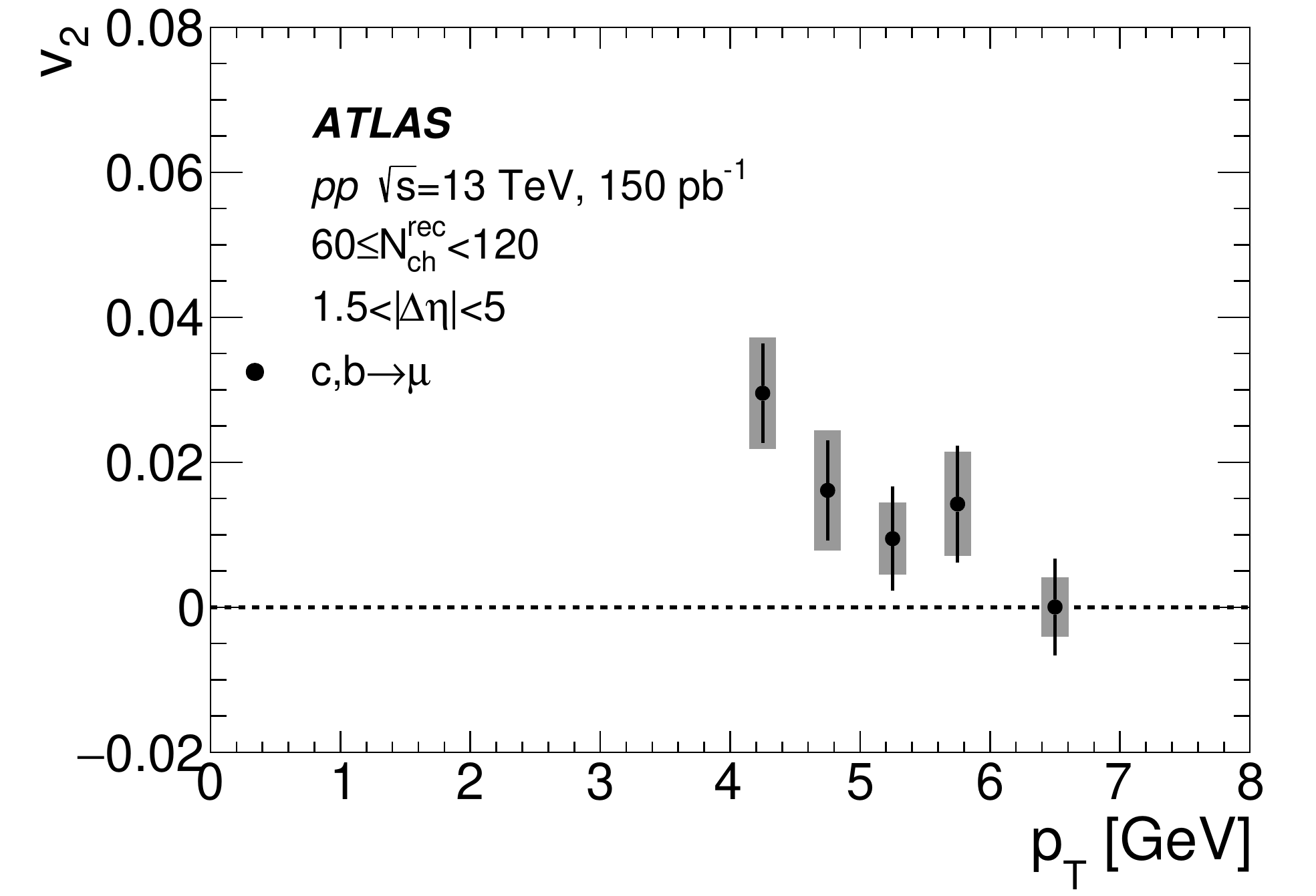}
\caption{\label{fig:v2_inclusive} Elliptic anisotropy coefficient
$v_{2}$ of inclusive heavy-flavor muons as a function of track
multiplicity \Nchrec for muons with transverse momentum
$4<\pt<6$~\GeV\ (left) and as a function of \pt for the
$60\leq\Nchrec <120$ multiplicity range (right). The vertical bars
and shaded bands represent statistical and systematic uncertainties,
respectively.}
\end{figure*}

\begin{figure*}[hbt]
\includegraphics[width=0.49\linewidth]{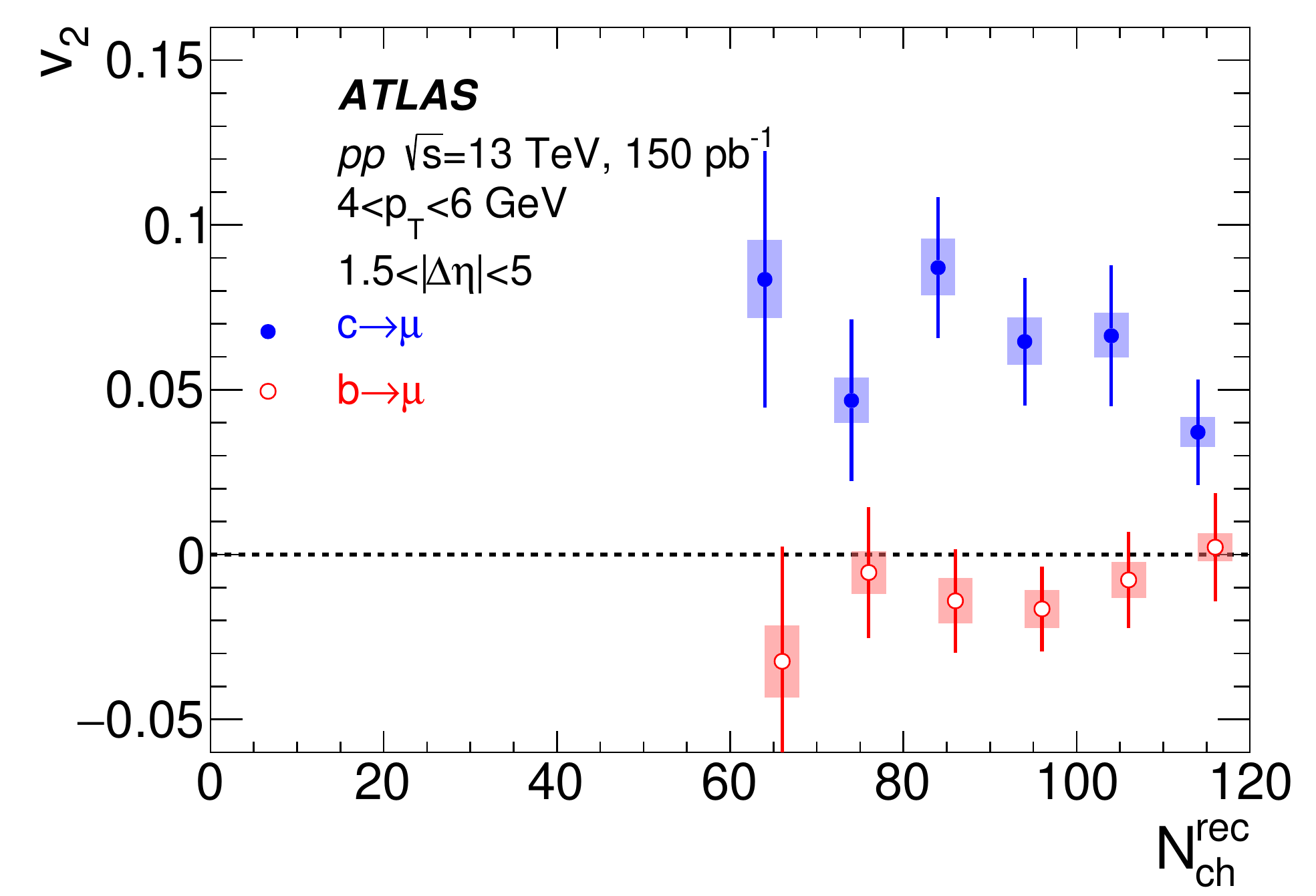}
\includegraphics[width=0.49\linewidth]{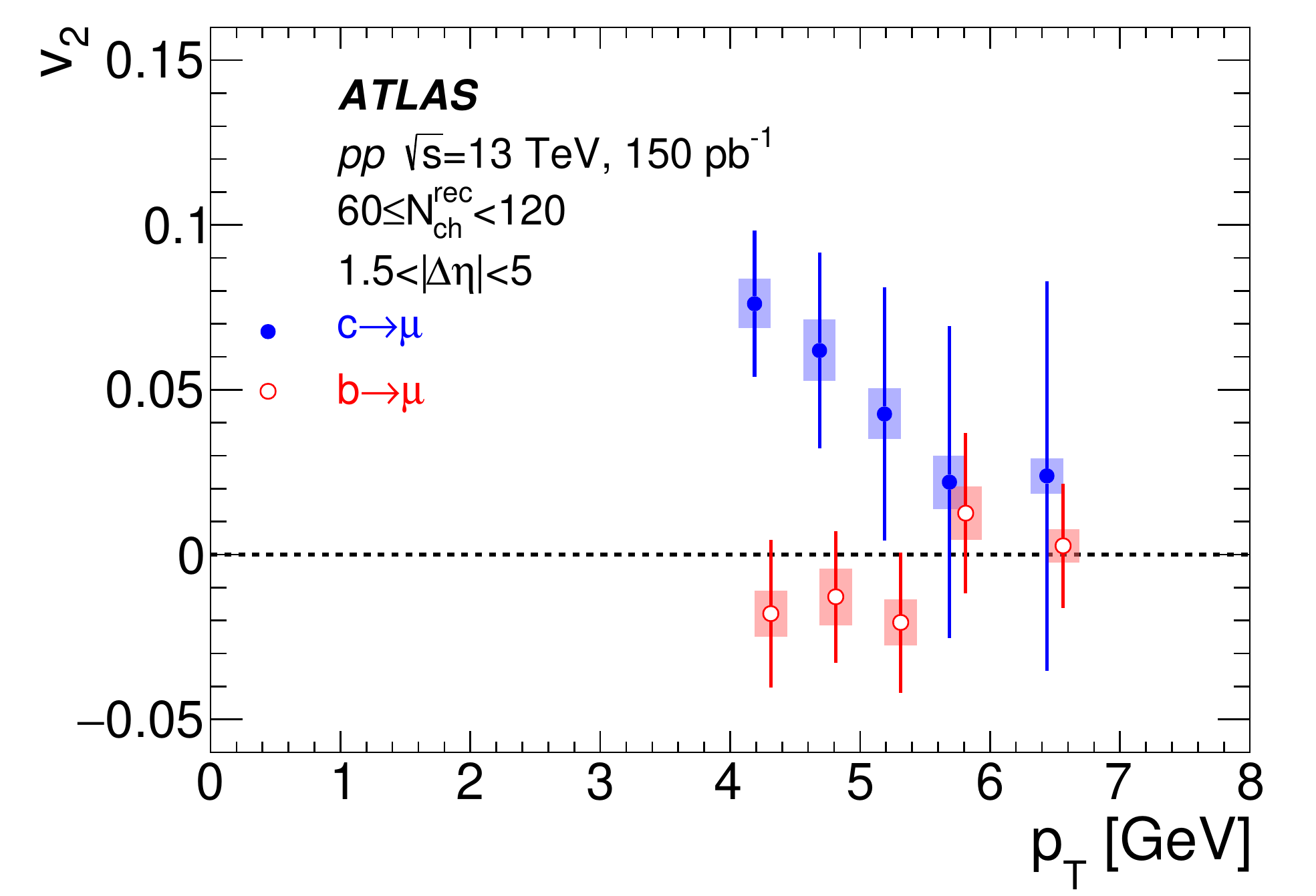}
\caption{\label{fig:v2_cb} Elliptic anisotropy coefficient $v_{2}$ of
muons from charm and bottom decays as a function of track
multiplicity \Nchrec for muons with transverse momentum
$4<\pt<6$~\GeV\ (left) and as a function of \pt for the
$60\leq\Nchrec <120$ multiplicity range (right). Data points are
shifted by $\pm1$ in \Nchrec and $\pm0.125$~\GeV\ in \pt for better
visibility. The vertical bars and shaded bands represent statistical
and systematic uncertainties, respectively.}
\end{figure*}
 
Figure~\ref{fig:v2_inclusive} shows the $v_2$ of inclusive
heavy-flavor muons, determined as $v^{\mu}_2(\pt^{\mu}) =
v_{2,2}(\pt^{\mu},\pt^{h})/v^{h}_{2}(\pt^{h})$, where
$v^{h}_{2}(\pt^{h})$ is taken from Ref.~\cite{HION-2016-01}. The
systematic uncertainty in the charged-hadron $v_2$ is included in the
total uncertainty, but is negligible compared with the other
uncertainties introduced in this measurement.  The $v_2$ value is
presented as a function of \Nchrec for $4<\pt<6$~\GeV\ (left) and as a
function of \pt for $60\leq\Nchrec< 120$ (right).
Within the uncertainties there is no clear \Nchrec dependence, but the value decreases as the heavy-flavor muon \pt increases from 4 to 7~\GeV.
 
Figure~\ref{fig:v2_cb} shows the $v_2$ values for muons from charm and
bottom decays separately, as a function of \Nchrec for
$4<\pt<6$~\GeV\ (left) and as a function of \pt for $60\leq\Nchrec<
120$ (right). The $v_2$ of muons from bottom decays is consistent with
zero in the entire \Nchrec range of the measurement and has no
discernible \pt dependence. In contrast, the $v_2$ of muons from charm
decays is nonzero at lower \pt but consistent with zero at higher
\pt within the sizeable uncertainties.  It also shows no significant \Nchrec dependence within the
uncertainties.

In summary, a measurement of elliptic flow coefficients for
heavy-flavor decay muons in \pp collisions at 13~\TeV\ is presented,
including a separation between charm and bottom contributions. The
measurement uses a dataset corresponding to an integrated luminosity
of 150~pb$^{-1}$ recorded by the ATLAS experiment at the LHC.  The
inclusive heavy-flavor muon $v_{2}$ values are not dependent on
\Nchrec in the range 60--120 and show a clear decrease with \pt from 4
to 7~\GeV.  The bottom-decay muons have $v_{2}$ values consistent with
zero within statistical and systematic uncertainties, while the
charm-decay muons have significant non-zero $v_{2}$ values.  These
results indicate that bottom quarks, unlike light and charm quarks, do
not participate in the collective behavior in high-multiplicity \pp
collisions.  There are theoretical calculations within a linearized
Boltzmann-Langevin transport framework for \PbPb collisions at
$\sqrt{s_{NN}}=5.02~\mathrm{TeV}$ predicting larger $v_{2}$ for D
meson than $v_2$ for B meson at $\pt<10~\mathrm{GeV}$ and similar
$v_2$ at $\pt>10~\mathrm{GeV}$~\cite{Ke:2018tsh}.  However, no such
calculations have been published for smaller systems including
high-multiplicity \pp events.  The results will provide fundamental
new input to the theoretical models which attempt to
describe heavy-quark transport and energy loss in these smallest
collision systems.
 
% The next lines are included from the .//acknowledgements/Acknowledgements.tex input file

We thank CERN for the very successful operation of the LHC, as well as the
support staff from our institutions without whom ATLAS could not be
operated efficiently.
 
We acknowledge the support of ANPCyT, Argentina; YerPhI, Armenia; ARC, Australia; BMWFW and FWF, Austria; ANAS, Azerbaijan; SSTC, Belarus; CNPq and FAPESP, Brazil; NSERC, NRC and CFI, Canada; CERN; CONICYT, Chile; CAS, MOST and NSFC, China; COLCIENCIAS, Colombia; MSMT CR, MPO CR and VSC CR, Czech Republic; DNRF and DNSRC, Denmark; IN2P3-CNRS and CEA-DRF/IRFU, France; SRNSFG, Georgia; BMBF, HGF and MPG, Germany; GSRT, Greece; RGC and Hong Kong SAR, China; ISF and Benoziyo Center, Israel; INFN, Italy; MEXT and JSPS, Japan; CNRST, Morocco; NWO, Netherlands; RCN, Norway; MNiSW and NCN, Poland; FCT, Portugal; MNE/IFA, Romania; MES of Russia and NRC KI, Russia Federation; JINR; MESTD, Serbia; MSSR, Slovakia; ARRS and MIZ\v{S}, Slovenia; DST/NRF, South Africa; MINECO, Spain; SRC and Wallenberg Foundation, Sweden; SERI, SNSF and Cantons of Bern and Geneva, Switzerland; MOST, Taiwan; TAEK, Turkey; STFC, United Kingdom; DOE and NSF, United States of America. In addition, individual groups and members have received support from BCKDF, CANARIE, Compute Canada and CRC, Canada; ERC, ERDF, Horizon 2020, Marie Sk{\l}odowska-Curie Actions and COST, European Union; Investissements d'Avenir Labex, Investissements d'Avenir Idex and ANR, France; DFG and AvH Foundation, Germany; Herakleitos, Thales and Aristeia programmes co-financed by EU-ESF and the Greek NSRF, Greece; BSF-NSF and GIF, Israel; CERCA Programme Generalitat de Catalunya and PROMETEO Programme Generalitat Valenciana, Spain; G\"{o}ran Gustafssons Stiftelse, Sweden; The Royal Society and Leverhulme Trust, United Kingdom.
 
The crucial computing support from all WLCG partners is acknowledged gratefully, in particular from CERN, the ATLAS Tier-1 facilities at TRIUMF (Canada), NDGF (Denmark, Norway, Sweden), CC-IN2P3 (France), KIT/GridKA (Germany), INFN-CNAF (Italy), NL-T1 (Netherlands), PIC (Spain), ASGC (Taiwan), RAL (UK) and BNL (USA), the Tier-2 facilities worldwide and large non-WLCG resource providers. Major contributors of computing resources are listed in Ref.~\cite{ATL-GEN-PUB-2016-002}.
 
% End of text imported from the .//acknowledgements/Acknowledgements.tex input file
 
\clearpage
 
\printbibliography

@article{Khachatryan:2010gv,
      author         = "Khachatryan, Vardan and others",
      title          = "{Observation of Long-Range Near-Side Angular Correlations
                        in Proton-Proton Collisions at the LHC}",
      collaboration  = "CMS",
      journal        = "JHEP",
      volume         = "09",
      year           = "2010",
      pages          = "091",
      doi            = "10.1007/JHEP09(2010)091",
      eprint         = "1009.4122",
      archivePrefix  = "arXiv",
      primaryClass   = "hep-ex",
      reportNumber   = "CMS-QCD-10-002, CERN-PH-EP-2010-031",
      SLACcitation   = "%%CITATION = ARXIV:1009.4122;%%"
}

@article{Ke:2018tsh,
      author         = "Ke, Weiyao and Xu, Yingru and Bass, Steffen A.",
      title          = "{Linearized Boltzmann-Langevin model for heavy quark
                        transport in hot and dense QCD matter}",
      journal        = "Phys. Rev.",
      volume         = "C98",
      year           = "2018",
      number         = "6",
      pages          = "064901",
      doi            = "10.1103/PhysRevC.98.064901",
      eprint         = "1806.08848",
      archivePrefix  = "arXiv",
      primaryClass   = "nucl-th",
      SLACcitation   = "%%CITATION = ARXIV:1806.08848;%%"
}

@Article{HION-2016-01,
    author         = "{ATLAS Collaboration}",
    title          = "{Measurements of long-range azimuthal anisotropies and associated Fourier coefficients for \(pp\) collisions at \(\sqrt{s}=5.02\) and \(13~\text{TeV}\) and \(p\)+Pb collisions at \(\sqrt{s_{\text{NN}}}=5.02\) TeV with the ATLAS detector}",
    journal        = "Phys. Rev. C",
    volume         = "96",
    year           = "2017",
    pages          = "024908",
    doi            = "10.1103/PhysRevC.96.024908",
    reportNumber   = "CERN-EP-2016-200",
    eprint         = "1609.06213",
    archivePrefix  = "arXiv",
    primaryClass   = "nucl-ex",
}

@Article{HION-2015-06,
      author         = "{ATLAS Collaboration}",
      title          = "{Measurement of the suppression and azimuthal anisotropy
                        of muons from heavy-flavor decays in Pb+Pb collisions at
                        $\sqrt{s_{\mathrm{NN}}} = 2.76$ TeV with the ATLAS
                        detector}",
      journal        = "Phys. Rev. C",
      volume         = "98",
      year           = "2018",
      pages          = "044905",
      doi            = "10.1103/PhysRevC.98.044905",
      eprint         = "1805.05220",
      archivePrefix  = "arXiv",
      primaryClass   = "nucl-ex",
      reportNumber   = "CERN-EP-2018-072",
}

@article{Cacciari:2012ny,
      author         = "Cacciari, Matteo and Frixione, Stefano and Houdeau,
                        Nicolas and Mangano, Michelangelo L. and Nason, Paolo and
                        Ridolfi, Giovanni",
      title          = "{Theoretical predictions for charm and bottom production
                        at the LHC}",
      journal        = "JHEP",
      volume         = "10",
      year           = "2012",
      pages          = "137",
      doi            = "10.1007/JHEP10(2012)137",
      eprint         = "1205.6344",
      archivePrefix  = "arXiv",
      primaryClass   = "hep-ph",
      reportNumber   = "CERN-PH-TH-2011-227",
      SLACcitation   = "%%CITATION = ARXIV:1205.6344;%%"
}

@article{Sjostrand:2007gs,
      author         = "Sj{\"{o}}strand, Torbj{\"{o}}rn and Mrenna, Stephen and Skands, Peter Z.",
      title          = "{A brief introduction to PYTHIA 8.1}",
      journal        = "Comput. Phys. Commun.",
      volume         = "178",
      year           = "2008",
      pages          = "852-867",
      doi            = "10.1016/j.cpc.2008.01.036",
      eprint         = "0710.3820",
      archivePrefix  = "arXiv",
      primaryClass   = "hep-ph",
      reportNumber   = "CERN-LCGAPP-2007-04, LU-TP-07-28,
                        FERMILAB-PUB-07-512-CD-T",
      SLACcitation   = "%%CITATION = ARXIV:0710.3820;%%"                        
}

@article{Agostinelli:2002hh,
      author         = "Agostinelli, S. and others",
      title          = "{GEANT4---a simulation toolkit}",
      collaboration  = "GEANT4",
      journal        = "Nucl. Instrum. Meth. A",
      volume         = "506",
      year           = "2003",
      pages          = "250-303",
      doi            = "10.1016/S0168-9002(03)01368-8",
      reportNumber   = "SLAC-PUB-9350, FERMILAB-PUB-03-339",
      SLACcitation   = "%%CITATION = NUIMA,A506,250;%%"  
}

@article{Ball:2012cx,
      author         = "Ball, Richard D. and others",
      title          = "{Parton distributions with LHC data}",
      journal        = "Nucl. Phys. B",
      volume         = "867",
      year           = "2013",
      pages          = "244-289",
      doi            = "10.1016/j.nuclphysb.2012.10.003",
      eprint         = "1207.1303",
      archivePrefix  = "arXiv",
      primaryClass   = "hep-ph",
      reportNumber   = "EDINBURGH-2012-08, IFUM-FT-997, FR-PHENO-2012-014,
                        RWTH-TTK-12-25, CERN-PH-TH-2012-037, SFB-CPP-12-47",
      SLACcitation   = "%%CITATION = ARXIV:1207.1303;%%"                                                                                                       
}

@Article{Adare:2008ae,
      author         = "{PHENIX Collaboration}",
      title          = "{Dihadron azimuthal correlations in Au$+$Au collisions at
                        $\sqrt{s_{NN}}=200~\mathrm{GeV}$}",
      journal        = "Phys. Rev. C",
      volume         = "78",
      year           = "2008",
      pages          = "014901",
      doi            = "10.1103/PhysRevC.78.014901",
      eprint         = "0801.4545",
      archivePrefix  = "arXiv",
      primaryClass   = "nucl-ex",
}

@article{Romatschke:2017ejr,
      author         = "Romatschke, Paul and Romatschke, Ulrike",
      title          = "{Relativistic Fluid Dynamics In and Out of Equilibrium --
                        Ten Years of Progress in Theory and Numerical Simulations
                        of Nuclear Collisions}",
      year           = "2017",
      eprint         = "1712.05815",
      archivePrefix  = "arXiv",
      primaryClass   = "nucl-th",
      SLACcitation   = "%%CITATION = ARXIV:1712.05815;%%"
}

@article{Heinz:2013th,
      author         = "Heinz, Ulrich and Snellings, Raimond",
      title          = "{Collective Flow and Viscosity in Relativistic Heavy-Ion
                        Collisions}",
      journal        = "Ann. Rev. Nucl. Part. Sci.",
      volume         = "63",
      year           = "2013",
      pages          = "123-151",
      doi            = "10.1146/annurev-nucl-102212-170540",
      eprint         = "1301.2826",
      archivePrefix  = "arXiv",
      primaryClass   = "nucl-th",
      SLACcitation   = "%%CITATION = ARXIV:1301.2826;%%"
}

@article{Batsouli:2002qf,
      author         = "Batsouli, S. and Kelly, S. and Gyulassy, M. and Nagle, J.
                        L.",
      title          = "{Does the charm flow at RHIC?}",
      journal        = "Phys. Lett. B",
      volume         = "557",
      year           = "2003",
      pages          = "26-32",
      doi            = "10.1016/S0370-2693(03)00175-8",
      eprint         = "nucl-th/0212068",
      archivePrefix  = "arXiv",
      SLACcitation   = "%%CITATION = NUCL-TH/0212068;%%"
}

@article{Adare:2006nq,
      author         = "{PHENIX Collaboration}",
      title          = "{Energy Loss and Flow of Heavy Quarks in Au+Au Collisions
                        at $\sqrt{s_{NN}}=200$ GeV}",
      journal        = "Phys. Rev. Lett.",
      volume         = "98",
      year           = "2007",
      pages          = "172301",
      doi            = "10.1103/PhysRevLett.98.172301",
      eprint         = "nucl-ex/0611018",
      archivePrefix  = "arXiv",
}

@article{Dong:2019byy,
      author         = "Dong, Xin and Lee, Yen-Jie and Rapp, Ralf",
      title          = "{Open Heavy-Flavor Production in Heavy-Ion Collisions}",
      year           = "2019",
      eprint         = "1903.07709",
      archivePrefix  = "arXiv",
      primaryClass   = "nucl-ex",
      SLACcitation   = "%%CITATION = ARXIV:1903.07709;%%"
}

@article{Nagle:2018nvi,
      author         = "Nagle, James L. and Zajc, William A.",
      title          = "{Small System Collectivity in Relativistic Hadronic and
                        Nuclear Collisions}",
      journal        = "Ann. Rev. Nucl. Part. Sci.",
      volume         = "68",
      year           = "2018",
      pages          = "211-235",
      doi            = "10.1146/annurev-nucl-101916-123209",
      eprint         = "1801.03477",
      archivePrefix  = "arXiv",
      primaryClass   = "nucl-ex",
      SLACcitation   = "%%CITATION = ARXIV:1801.03477;%%"
}

@article{Weller:2017tsr,
      author         = "Weller, Ryan D. and Romatschke, Paul",
      title          = "{One fluid to rule them all: Viscous hydrodynamic
                        description of event-by-event central p+p, p+Pb and Pb+Pb
                        collisions at $\sqrt{s}=5.02$ TeV}",
      journal        = "Phys. Lett. B",
      volume         = "774",
      year           = "2017",
      pages          = "351-356",
      doi            = "10.1016/j.physletb.2017.09.077",
      eprint         = "1701.07145",
      archivePrefix  = "arXiv",
      primaryClass   = "nucl-th",
      SLACcitation   = "%%CITATION = ARXIV:1701.07145;%%"
}

@Article{Sirunyan:2018toe,
      author         = "{CMS Collaboration}",
      title          = "{Elliptic Flow of Charm and Strange Hadrons in
                        High-Multiplicity p+Pb Collisions at
                        $\sqrt{s_{_\mathrm{NN}}} =$ 8.16 TeV}",
      journal        = "Phys. Rev. Lett.",
      volume         = "121",
      year           = "2018",
      pages          = "082301",
      doi            = "10.1103/PhysRevLett.121.082301",
      eprint         = "1804.09767",
      archivePrefix  = "arXiv",
      primaryClass   = "hep-ex",
      reportNumber   = "CMS-HIN-17-003, CERN-EP-2018-076",
}

@article{Moore:2004tg,
      author         = "Moore, Guy D. and Teaney, Derek",
      title          = "{How much do heavy quarks thermalize in a heavy ion
                        collision?}",
      journal        = "Phys. Rev. C",
      volume         = "71",
      year           = "2005",
      pages          = "064904",
      doi            = "10.1103/PhysRevC.71.064904",
      eprint         = "hep-ph/0412346",
      archivePrefix  = "arXiv",
      SLACcitation   = "%%CITATION = HEP-PH/0412346;%%"
}

@article{Adam:2016ich,
      author         = "{ALICE Collaboration}",
      title          = "{$D$-meson production in $p$-Pb collisions at
                        $\sqrt{s_\mathrm{NN}}=$5.02 TeV and in pp collisions at
                        $\sqrt{s}=$7 TeV}",
      journal        = "Phys. Rev. C",
      volume         = "94",
      year           = "2016",
      pages          = "054908",
      doi            = "10.1103/PhysRevC.94.054908",
      eprint         = "1605.07569",
      archivePrefix  = "arXiv",
      primaryClass   = "nucl-ex",
      reportNumber   = "CERN-EP-2016-127",
}

@article{Acharya:2018dxy,
      author         = "{ALICE Collaboration}",
      title          = "{Azimuthal Anisotropy of Heavy-Flavor Decay Electrons in
                        $p$-Pb Collisions at $ \sqrt{s_\mathrm{NN}}$ = 5.02 TeV}",
%      collaboration  = "ALICE",
      journal        = "Phys. Rev. Lett.",
      volume         = "122",
      year           = "2019",
      number         = "7",
      pages          = "072301",
      doi            = "10.1103/PhysRevLett.122.072301",
      eprint         = "1805.04367",
      archivePrefix  = "arXiv",
      primaryClass   = "nucl-ex",
      reportNumber   = "CERN-EP-2018-119",
      SLACcitation   = "%%CITATION = ARXIV:1805.04367;%%"
}

@article{Lim:2019cys,
      author         = "Lim, S. H. and Hu, Q. and Belmont, R. and Hill, K. K. and
                        Nagle, J. L. and Perepelitsa, D. V.",
      title          = "{Examination of flow and nonflow factorization methods in
                        small collision systems}",
      journal        = "Phys. Rev.",
      volume         = "C100",
      year           = "2019",
      number         = "2",
      pages          = "024908",
      doi            = "10.1103/PhysRevC.100.024908",
      eprint         = "1902.11290",
      archivePrefix  = "arXiv",
      primaryClass   = "nucl-th",
      SLACcitation   = "%%CITATION = ARXIV:1902.11290;%%"
}

@article{Dokshitzer:2001zm,
      author         = "Dokshitzer, Yuri L. and Kharzeev, D. E.",
      title          = "{Heavy-quark colorimetry of QCD matter}",
      journal        = "Phys. Lett. B",
      volume         = "519",
      year           = "2001",
      pages          = "199-206",
      doi            = "10.1016/S0370-2693(01)01130-3",
      eprint         = "hep-ph/0106202",
      archivePrefix  = "arXiv",
      reportNumber   = "LPT-ORSAY-01-58, BNL-NT-01-9",
      SLACcitation   = "%%CITATION = HEP-PH/0106202;%%"
}

@article{Sharma:2009hn,
      author         = "Sharma, Rishi and Vitev, Ivan and Zhang, Ben-Wei",
      title          = "{Light-cone wave function approach to open heavy flavor
                        dynamics in QCD matter}",
      journal        = "Phys. Rev. C",
      volume         = "80",
      year           = "2009",
      pages          = "054902",
      doi            = "10.1103/PhysRevC.80.054902",
      eprint         = "0904.0032",
      archivePrefix  = "arXiv",
      primaryClass   = "hep-ph",
      SLACcitation   = "%%CITATION = ARXIV:0904.0032;%%"
}

@Article{PERF-2007-01,
    author         = "{ATLAS Collaboration}",
    title          = "{The ATLAS Experiment at the CERN Large Hadron Collider}",
    journal        = "JINST",
    volume         = "3",
    year           = "2008",
    pages          = "S08003",
    doi            = "10.1088/1748-0221/3/08/S08003",
    primaryClass   = "hep-ex",
}

@Article{SOFT-2010-01,
    author         = "{ATLAS Collaboration}",
    title          = "{The ATLAS Simulation Infrastructure}",
    journal        = "Eur. Phys. J. C",
    volume         = "70",
    year           = "2010",
    pages          = "823",
    doi            = "10.1140/epjc/s10052-010-1429-9",
    eprint         = "1005.4568",
    archivePrefix  = "arXiv",
    primaryClass   = "physics.ins-det",
}

@Article{HION-2011-01,
    author         = "{ATLAS Collaboration}",
    title          = "{Measurement of the azimuthal anisotropy for charged particle production in \(\sqrt{s_{\text{NN}}}=2.76\,\text{TeV}\) lead--lead collisions with the ATLAS detector}",
    journal        = "Phys. Rev. C",
    volume         = "86",
    year           = "2012",
    pages          = "014907",
    doi            = "10.1103/PhysRevC.86.014907",
    reportNumber   = "CERN-PH-EP-2012-035",
    eprint         = "1203.3087",
    archivePrefix  = "arXiv",
    primaryClass   = "hep-ex",
}

@Article{STDM-2011-07,
    author         = "{ATLAS Collaboration}",
    title          = "{Measurements of the electron and muon inclusive cross-sections in proton--proton collisions at \(\sqrt{s} = 7\,\text{TeV}\) with the ATLAS detector}",
    journal        = "Phys. Lett. B",
    volume         = "707",
    year           = "2012",
    pages          = "438",
    doi            = "10.1016/j.physletb.2011.12.054",
    reportNumber   = "CERN-PH-EP-2011-108",
    eprint         = "1109.0525",
    archivePrefix  = "arXiv",
    primaryClass   = "hep-ex",
}

@Article{HION-2015-09,
    author         = "{ATLAS Collaboration}",
    title          = "{Observation of Long-Range Elliptic Azimuthal Anisotropies in \(\sqrt{s} = 13\) and \(2.76\,\text{TeV}\) \(pp\) Collisions with the ATLAS Detector}",
    journal        = "Phys. Rev. Lett.",
    volume         = "116",
    year           = "2016",
    pages          = "172301",
    doi            = "10.1103/PhysRevLett.116.172301",
    reportNumber   = "CERN-PH-EP-2015-251",
    eprint         = "1509.04776",
    archivePrefix  = "arXiv",
    primaryClass   = "hep-ex",
}

@Article{PERF-2015-10,
    author         = "{ATLAS Collaboration}",
    title          = "{Muon reconstruction performance of the ATLAS detector in proton--proton collision data at \(\sqrt{s} = 13\,\text{TeV}\)}",
    journal        = "Eur. Phys. J. C",
    volume         = "76",
    year           = "2016",
    pages          = "292",
    doi            = "10.1140/epjc/s10052-016-4120-y",
    reportNumber   = "CERN-EP-2016-033",
    eprint         = "1603.05598",
    archivePrefix  = "arXiv",
    primaryClass   = "hep-ex",
}

@Article{TRIG-2016-01,
    author         = "{ATLAS Collaboration}",
    title          = "{Performance of the ATLAS trigger system in 2015}",
    journal        = "Eur. Phys. J. C",
    volume         = "77",
    year           = "2017",
    pages          = "317",
    doi            = "10.1140/epjc/s10052-017-4852-3",
    reportNumber   = "CERN-EP-2016-241",
    eprint         = "1611.09661",
    archivePrefix  = "arXiv",
    primaryClass   = "hep-ex",
}

@Booklet{ATL-GEN-PUB-2016-002,
    author         = "{ATLAS Collaboration}",
    title          = "{ATLAS Computing Acknowledgements}",
    howpublished   = "{ATL-GEN-PUB-2016-002}",
    url            = "https://cds.cern.ch/record/2202407",
}

@Booklet{ATL-PHYS-PUB-2014-021,
    author         = "{ATLAS Collaboration}",
    title          = "{ATLAS Pythia~8 tunes to \(7~\text{TeV}\) data}",
    howpublished   = "{ATL-PHYS-PUB-2014-021}",
    url            = "https://cds.cern.ch/record/1966419",
    year           = "2014",
}

@Booklet{ATL-PHYS-PUB-2015-018,
    author         = "{ATLAS Collaboration}",
    title          = "{Track Reconstruction Performance of the ATLAS Inner Detector at  \(\sqrt{s} = 13~\text{TeV}\)}",
    howpublished   = "{ATL-PHYS-PUB-2015-018}",
    url            = "https://cds.cern.ch/record/2037683",
    year           = "2015",
}

\clearpage % ATLAS Collaboration author list
% Reference date of HION-2018-23 is 2019-06-06
% Author list last updated on date 06-MAR-20
% Data extracted on 06-Mar-2020 for paper reference HION-2018-23
% at 3:09pm
 
\begin{flushleft}
{\Large The ATLAS Collaboration}

\bigskip

G.~Aad$^\textrm{\scriptsize 102}$,    
B.~Abbott$^\textrm{\scriptsize 129}$,    
D.C.~Abbott$^\textrm{\scriptsize 103}$,    
A.~Abed~Abud$^\textrm{\scriptsize 71a,71b}$,    
K.~Abeling$^\textrm{\scriptsize 53}$,    
D.K.~Abhayasinghe$^\textrm{\scriptsize 94}$,    
S.H.~Abidi$^\textrm{\scriptsize 167}$,    
O.S.~AbouZeid$^\textrm{\scriptsize 40}$,    
N.L.~Abraham$^\textrm{\scriptsize 156}$,    
H.~Abramowicz$^\textrm{\scriptsize 161}$
% ...
\bigskip
\\

% ...
$^{40}$Niels Bohr Institute, University of Copenhagen, Copenhagen; Denmark.\\
% ...
$^{53}$II. Physikalisches Institut, Georg-August-Universit\"{a}t G\"ottingen, G\"ottingen; Germany.\\
% ...
$^{71}$$^{(a)}$INFN Sezione di Pavia;$^{(b)}$Dipartimento di Fisica, Universit\`a di Pavia, Pavia; Italy.\\
% ...
$^{94}$Department of Physics, Royal Holloway University of London, Egham; United Kingdom.\\
% ...
$^{102}$CPPM, Aix-Marseille Universit\'e, CNRS/IN2P3, Marseille; France.\\
$^{103}$Department of Physics, University of Massachusetts, Amherst MA; United States of America.\\
% ...
$^{129}$Homer L. Dodge Department of Physics and Astronomy, University of Oklahoma, Norman OK; United States of America.\\
% ...
$^{156}$Department of Physics and Astronomy, University of Sussex, Brighton; United Kingdom.\\
% ...
$^{161}$Raymond and Beverly Sackler School of Physics and Astronomy, Tel Aviv University, Tel Aviv; Israel.\\
% ...
$^{167}$Department of Physics, University of Toronto, Toronto ON; Canada.\\
% ...
$^{a}$ Also at Borough of Manhattan Community College, City University of New York, New York NY; United States of America.\\
$^{b}$ Also at CERN, Geneva; Switzerland.\\
% ...

\end{flushleft}

% Created with Glance <Atlas.Glance@cern.ch>

\end{document}